# Deep ConvLSTM with self-attention for human activity decoding using wearable sensors

Satya P. Singh, Sukrit Gupta, Madan Kumar Sharma, Aimé Lay-Ekuakille*, Deepak Gangwar

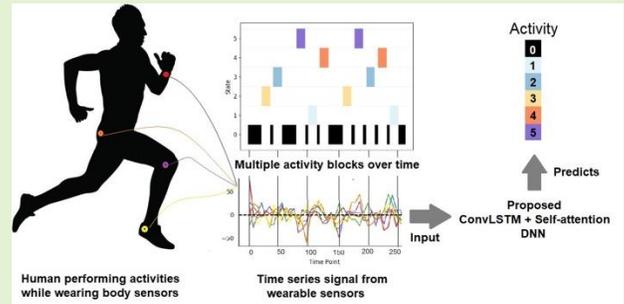

*Abstract*—**Decoding human activity accurately from wearable sensors can aid in applications related to healthcare and context awareness. The present approaches in this domain use recurrent and/or convolutional models to capture the spatio-temporal features from time-series data from multiple sensors. We propose a deep neural network architecture that not only captures the spatio-temporal features of multiple sensor time-series data but also selects, learns important time points by utilizing a self-attention mechanism. We show the validity of the proposed approach across different data sampling strategies on six public datasets and demonstrate that the self-attention mechanism gave a significant improvement in performance over deep networks using a combination of recurrent and convolution networks. We also show that the proposed approach gave a statistically significant performance enhancement over previous state-of-the-art methods for the tested datasets. The proposed methods open avenues for better decoding of human activity from multiple body sensors over extended periods of time. The code implementation for the proposed model is available at**
*https://github.com/isukrit/encodingHumanActivity*

*Index Terms*—**CNN, LSTM, Self-attention, wearable sensors, human activities**

## I. INTRODUCTION

DECODING human activity has varied emerging applications for human wellbeing that include monitoring health [1], living condition [2], and daily activities[3], [4]. The most popular methods used to monitor human activities are based on computer vision and wearable sensors. For computer vision-related methods, typically a camera is used to track and record human activity [5]. This requires adjusting the camera at different angles and using deep learning on images acquired from the camera that makes it difficult to decode the human activity being performed. On the other hand, wearable sensors are usually tiny microelectronics or biochemical-based devices that can be placed on different parts of the body as needed or can be put into the pocket of the subject [6]. Furthermore, with advances in wireless technologies, it is becoming possible to manufacture portable, low power, and highly efficient wearable sensor units [7], [8]. A wearable sensor unit consists of i) input data sensors (e.g. accelerometer, gyroscope) to extract the physiological or movement data, ii) signal transfer unit (wired or wireless medium) to send the collected data to the remote unit, and iii) signal processing and data analysis unit for extracting the relevant features and information [9]. Sometimes, the signals extracted from input sensors are noisy and low power, they require filtering and amplification before feature extraction and classification can be performed on them. For example, data collected using accelerometers, gyroscopes, and magnetometers can be affected by a variety of artifacts such as spurious or electronic noise resulting in abnormality or inconsistent values in the data. These anomalies can be remove using filtering or screening process [10], [11].

Many researchers use handcrafted features from the time-series data generated from wearable sensors placed on the different body parts [8], [12]–[15]. Popular handcrafted features include statistical features (mean, variance) and frequency domain features such as Fourier transform [16]. However, handcrafted features are highly data-dependent and they are much generalizable across application domains [17]. Furthermore, feature extraction is time-consuming and often only a limited number of features can be generated from a

Satya P. Singh is with Nanyang Technological University, Singapore(e-mail: satya@ntu.edu.sg).

Sukrit Gupta is with the School of Computer Science & Engineering, Nanyang Technological University, Singapore (e-mail: sukrit001@ntu.edu.sg)

Deepak Gangwar is with the Department of Electronics and Communication Engineering, Bharati Vidhyapeeth's Colleg(e-mail: er.deepakgangwar@gmail.com).

Madan Kumar Sharma is also with the Galgotias College of Engineering and Technology, India (e-mail: madansharma12@gmail.com)

*Corresponding author: Aimé Lay-Ekuakille is with the Department of Innovation Engineering University of Salento, Lecce, Italy (e-mail: aime.lay.ekuakille@unisalento.it)

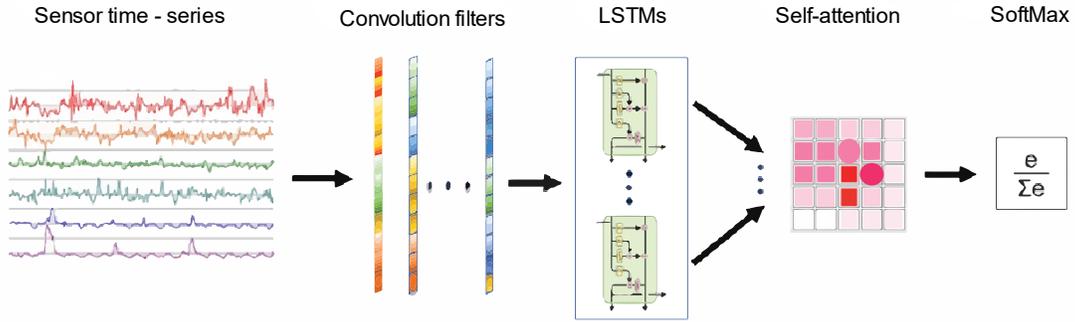

Figure 1 Framework of proposed model illustrating different steps dedicated to the algorithm with a special focus on the results connected to the self-attention which is pointed-out by two sub-processings included in the LSTMs unit.

given dataset. Handcrafted features are used to classify the samples with supervised machine learning algorithms such as support vector machine (SVM), AdaBoost, decision tree, feed-forward neural network (FFN), etc. Recent approaches, overcome the limitations of handcrafted features by exploring deep neural network (DNN) approaches for automatic feature learning from the complex datastructures [8], [12], [18]–[20]. DNNs can be given raw input features as input without the need for feature handcrafting. Therefore, given enough training samples, DNNs have better generalization capability compared to methods that use handcrafted features.

In addition, the availability of high-end GPUs allows building deeper models to learn descriptive features from complex data. In this regard, convolutional neural networks (CNN) have gained a lot of attention over the years[8], [21]. CNNs extract features within itself using convolutional operations and are domain-independent with generalization capabilities[22]. Despite their performance improvements, CNNs are computationally expensive and require a large number of training samples.

This has been partly mitigated by advancement in graphical processing units (GPUs), data augmentation techniques, and efficient neural network architectures making CNNs popular for human activity recognition using wearable sensors data. Data augmentation (by techniques such as rotation, permutation, time-warping, scaling, jittering, etc.) from data of wearable sensors improves the generalization ability and prevent overfitting in the DNN [23]. While CNNs capture the spatial domain (signals which can be represented by a point, line, or polygon and can be store in a database) of the sensor data and give reasonable performance for simple human activities such as walking, jogging, and eating [24], they are unable to capture complex activities that need analysis of temporal characteristics of the wearable sensor time-series signal [25]. Temporal characteristics /features are those features which are associated with or change with time. This is exacerbated by the fact that the time series data from wearable sensors is noisy and that all the sensors may not be in use all the time. Therefore, recurrent neural networks (RNNs) that can also give importance to temporal information from time series data are used in activity recognition from wearable sensors [26].

RNN are special types of neural network which are specially designed to tackle time-dependent (temporal) input sequences such as speech recognition, machine translation, and sequence to sequence learning. However, an issue with initial RNN models was with the flow of gradient, whereby the RNN suffered from the vanishing or exploding gradient problem that made them hard to train. This was solved by the advent of long short-term memory (LSTM) networks that add additional gates for information flow between different time points. LSTMs are very popular in natural language processing domain where they are used for word prediction, language translation, etc. Multiple studies in the literature propose a combination of CNN and LSTM [27]–[31], by taking advantage of CNN for capturing the spatial domain information and LSTMs for temporal domain information. For example, Ordóñez et al. [31] proposed Deep ConvLSTM where they use 4 consecutive CNN layers to extract local contextual features (abstract information) from the input time-series data. Further, these features were fed to LSTM units for extracting long temporal features (temporal dynamics).

While CNN and LSTM effectively capture spatio-temporal information, there is a need to focus on specific information from the embeddings generated by the CNN and LSTM combination. This is because the inputs in the form of time series from multiple sensors have latent relationships with each other that can be used for decoding human activity. This was not considered by previous studies in the area, and can be performed using the attention mechanism [32], [33] that learns the hidden relation between input variables. In this paper, we focus on the self-attention mechanism, although other attention mechanisms can also be used. The self-attention weighs the embedding of the input using a 2D matrix such that each row of the matrix caters to a different part of the sentence. Together with CNN and LSTM, we show that the self-attention mechanism leads to a statistically significant improvement in accuracy for six benchmark datasets.

## II. Proposed model for selecting sensors and time points

The architecture for the proposed model is illustrated in Fig. 1. The input for the proposed model is the time series in a time window of size $T$ from $N$ sensors. Let the input time series be $x = (x(t))_{t=1}^{T}$ where $x(t) \in \mathbb{R}^N$ is the input at time point $t$. It consists of three sub-modules: (1) an embedding layer consisting multiple 1-dimensional convolution filters to learn embeddings (local contextual feature or abstract information) from the inputs of wearable sensors; (2) an encoder consisting of one or more long short term memory (LSTM) layers (to extract long temporal features or temporal dynamics from the abstract information in the preceding CNN layer); and (3) an attention module consisting of a self-attention layer to discover and learn relationships between the time points in the input window. We add a SoftMax classification layer on top of these sub-modules.

### A. Embedding layer

The idea behind adding the embedding layer is to learn a representation of the wearable sensors using $K$ 1-dimensional spatial convolution operations. Let the filter weights for $k$th filter be denoted by $w_k$ and $h_c(t) = (h_{c,k}(t))_{k=1}^K$ be the output of the convolution layer where $h_{c,k}(t)$ is given by:

$$h_{c,k}(t) = x(t) \circledast w_k \quad (1)$$

where $\circledast$ denotes the convolution operation in the spatial domain, $w_k \in \mathbb{R}^N$ and output $h_c(t) \in R^{1 \times K}$ where $K$ is the number of filters.

### B. LSTM Encoder

We capture dependencies among time points using an RNN encoder architecture that learns temporal information from sensor time-series data. The encoder can consist of one or more LSTM layers. For one LSTM layer, the encoder output $h_e(t) \in \mathbb{R}^{1 \times E}$, where $E$ is the number of hidden units of the encoder LSTM, is given by:

$$h_e(t) = lstm(h_c(t), h_e(t-1)) \quad (2)$$

### C. Self-Attention Layer

After leveraging both local contextual features and temporal dynamics fusing CNN layer and LSTM units from the input time series, we use self-attention layer to learn weight coefficients which capture the relationship between time points in input data samples. The attention layer aims to learn the relationship between input time points from the sensor time-series data that aid in determining the state label. We feed the concatenated outputs of the encoder, $h_e \in \mathbb{R}^{T \times E}$, for different time points to the attention layer:

$$h_e = [h_e(t)]_{t=1}^T \quad (3)$$

The $h_e$ matrix captures the representation of the input sensor data over multiple time points $T$. The attention score, $S \in \mathbb{R}^{F \times E}$, for the sample $x$ is then given by:

$$\alpha = softmax(V_{att} tanh(U_{att} h'_e)) \quad (4)$$

$$S = \alpha \cdot h_e \quad (5)$$

where $h'_e$ is the transpose of $h_e$, $U_{att} \in \mathbb{R}^{D \times E}$ and $V_{att} \in \mathbb{R}^{F \times D}$ are weight matrices forming the attention module, $D$ represents the attention length, $F$ represents the length of the output and the dot product ($\cdot$) is taken in the spatial domain.

The $\alpha \in \mathbb{R}^{F \times T}$ has values in the range $(0, 1)$ and is used to give different weights to each $h_e(t)$ from $h_e$, i.e. embedding $h_e(t)$ for each time point is weighted in the attention score $S$.

### D. SoftMax Layer

The flattened attention score vector $s \in \mathbb{R}^{(F*E) \times 1}$ is used as input to the output softmax layer. The softmax layer output $y^*$ is given by:

$$y^* = softmax(Ws + b) \quad (6)$$

where $W \in \mathbb{R}^{C \times (F*E)}$ and $b \in \mathbb{R}^C$ denote the weight matrix and bias vector of the softmax layer and $C$ denotes the number of possible output states. The output state label is determined by the state (i.e., the neuron) receiving the maximum output.

### E. Learning

The cross-entropy cost, $J$, is minimized to train the proposed model, such that $J$ is given by:

$$J = -E_x[y \log(p(y^*))] \quad (7)$$

where $E_x$ denotes the expectation over inputs $x$ and $y$ denotes the state label. We use the Adam optimizer to minimize the cross-entropy cost.

## III. EXPERIMENTAL SETUP

### A. Dataset description

There are many datasets are available for human activity recognition using wearable sensors. However, most of them are collected with different sampling rate, number of sensors, sensor placement, and number of recorded activities. The proposed approach and existing approaches presented in the literature were experimented and validated using six benchmark wearable sensor datasets which include mobile health (MHEALTH) [10], [11], UTD Multimodal Human Action Dataset (UTD-MHAD) [34], USC human activity dataset (USC-HAD)[35], Wearable Human Activity Recognition Folder (WHARF)[36], and Wireless Sensor Data Mining (WISDM)[37]. These datasets are prepared with different sampling rates, number of sensors, and recorded number of activities. Also, while some of these datasets are balanced and some of them are highly imbalanced. Details about data can be found at the respective website and also summarized in Table I. The following sections discuss the aforementioned datasets briefly:

MHEALTH: Body signals and vital signs (rate of turn, acceleration, Magnetic field orientations) were recorded from sensors placed on left ankle, chest, and right wrist of 12 volunteers. All activities were recorded using a sampling rate of 50 HZ.

UTD-MHAD: The UTD-MHAD is a kind of multimodal human action dataset prepared by the fusion of 16-bit depth with a resolution of 320x240 pixels and low cost wireless inertial data sensors with a sampling rate of 50 Hz. There is a total of 27 actions recorded using wearable sensors from 8 subjects. For action 1-21, the inertial sensor is placed on the subject's right wrist and for 22-27, the sensor was placed on the subject's right thigh.

Table I A brief description of the dataset used in experimental and validation of the proposed and similar work found in the literature.

| Dataset | Sensors type and location | Sampling freq. | # Volunteers | Sensor placement | #Activities |
|---|---|---|---|---|---|
| MHEALTH [10] | accelerometer, gyroscope, and magnetometer | 50 Hz | 10 | Sensors were placed on subject's chest, right wrist and left ankle | 12 physical activities such as standing, sitting, lying, walking, climbing, cycling, jogging, running jumping. |
| USC-HAD [35] | Motion node (accelerometer, gyroscope, and magnetometer) | 100Hz | 15 | Motion node was packed into a mobile phone and attached to the subject's right hip | 10 physical activities such as walking, jumping, standing, sleeping, elevator op/down. |
| UTD-MHAD1 [34] | Wearable inertial sensors which capture 3-axis acceleration, 3-axis angular velocity and 3-axis magnetic strength | 50 Hz | 8 | Wearable inertial sensor on right wrist | 21 physical activities such as swipe, wave, throw, clap, drawing, bowling, boxing, tennis, push, knock. |
| UTD-MHAD2 [34] | --do-- | 50 Hz | 8 | Wearable inertial sensor on right thigh | 6 physical activities such as jogging, walking, sit, stand, foot forward. |
| WHARF [36] | single wrist-worn tri-axial accelerometer | 32 | 17 | volunteers with a wrist-mounted tri-axial accelerometer | 14 mechanical and location independent activities such as toileting, transferring, eating, sit/stand, climbing stairs, telephone, transpiration. |
| WISDM [37] | Phone based accelerometer | 20 | 29 | Smart phone in pants leg pocket | 6 daily activities such as walking, jogging, stairs, sitting, and standing. |

USC-HAD: USC-HAD dataset contains data related to 12 daily activities from 14 subjects. The data were recorded using a motion node firmly placed on the subject's right hip and the subjects were asked to perform their own style (such as walking stairs, walking forward, etc.). The motion node consists of an accelerometer (3-axis), gyroscope (3-axis), and magnetometer (3-axis). The maximum sampling rate was kept to 100 Hz.

WHARF: WHARF data were recorded using a single 3-axis accelerometer place on the subject's wrist. The data was collected from 17 subjects performing 14 human activities (such as brushing teeth, comping hair, feeding, etc.) The data is composed of over 1000 recordings with a sampling rate of 30 H.

WISDM: The dataset is a collection of raw time-series data generated from the accelerometer in the mobile phone. The device was placed on the waist of the subjects while performing daily activities (such as walking, jogging, etc.) The dataset was sampled at 20 Hz and contains 109827 examples.

Since the collection of datasets involved human subjects, appropriate permissions, consent, and approvals to perform activities were taken and can be found on the respective repositories.

B. Sample generation process

The first step in experimenting with the wearable sensor data is sample generation from raw time series from different sensors. We use a temporal window of fixed size to split the entire time series into equal parts. In cases where the length of the last data sample is not equal to other data samples, it is omitted. There are several techniques to use temporal windows for data split. The most commonly used window in the literature is the semi non-overlapping temporal window (SNOW) in which a fixed-size window is applied on the input data sequence to generate data for training and testing samples. However, this process is highly biased because there is a 50% overlap between sliding subsequent windows. To avoid this bias, there is another process called fully non-overlapping temporal window (FNOW). In this process, there is no overlap between the subsequent sliding windows. As obvious, this will generate a small number of data samples which may not fulfill the essential requirement (large data samples) for any deep learning models. In order to increase the sample size while keeping the data unbiased, Jorado et al. [38] have suggested the leave one trial out (LOTO)[1] process. A trial is defined as a single activity (raw signal) performed by the subject. Similar to [38], we ensure that during data split for training, validation and testing groups, the trial signal (raw time series from wearable sensors) from one data group must not be mixed with other groups during 10-fold cross validation. Further, we generated data samples using semi non-overlapping temporal window (SNOW). By doing so, we generated high number of data samples which is the necessity for any deep learning model without introducing any bias.

C. Evaluation protocols

The performance of classifiers employed in human activity recognition can be measured using several performance measurement indices such as accuracy, precision, recall, F1-score etc. In the current work, we choose three performance indices i.e. accuracy, recall, and F1-score. Accuracy is simply the ratio of correctly predicted samples to the total number of samples. Accuracy is the recommended metric to measure the classification performance, if the data is balanced. We define accuracy, recall, and F1-score as follows:

---

[1] https://github.com/arturjordao/WearableSensorData

Table II Performance of proposed (Accuracy%± standard deviation) approach using different sample generation methods.

| Sample generation | MHEALTH | USC-HAD | UTD-MHAD1 | UTD-MHAD2 | WHARF | WISDM |
|---|---|---|---|---|---|---|
| SNOW | 99.75±0.43 | 94.06±0.76 | 65.50±3.28 | 91.40±3.47 | 87.29±2.50 | 99.26±0.24 |
| FNOW | 99.84±0.33 | 91.40±1.02 | 53.90±3.71 | 86.29±4.32 | 80.05±4.06 | 98.27±0.47 |
| LOTO | 94.86±7.65 | 90.88±1.47 | 58.02±2.29 | 89.84±2.97 | 82.39±3.93 | 90.41±5.59 |

Table III Comparative analysis of proposed approach with baseline approach (deep ConvLSTM) using acc. (Accuracy%± std.), recall (recall%± std.), and F1-score(F1-score %± std.) for six benchmark datasets.

| Performance | MHEALTH | USC-HAD | UTD-MHAD1 | UTD-MHAD2 | WHARF | WISDM |
|---|---|---|---|---|---|---|
| | Proposed approach, ConvLSTM with self-attention | | | | | |
| Acc. | 94.86±7.65 | 90.88±1.47 | 58.02±2.29 | 89.84±2.97 | 82.39±3.93 | 90.41±5.59 |
| Recall | 94.35±8.76 | 87.69±1.53 | 55.15±4.26 | 73.27±10.41 | 78.36±3.42 | 88.19±6.99 |
| F1-score | 93.74±10.04 | 86.22±1.52 | 54.41±3.00 | 74.06±9.17 | 77.73±3.32 | 86.88±7.49 |
| | Baseline ConvLSTM | | | | | |
| Acc. | 93.80±3.87 | 87.05±3.2 | 43.92±3.71 | 87.06±3.74 | 77.80±4.07 | 89.83±7.04 |
| Recall | 92.87±3.89 | 83.2±3.93 | 42.75±4.12 | 86.26±4.63 | 71.83±5.31 | 88.38±6.8 |
| F1-score | 92.35±4.38 | 81.79±4.31 | 42.38±4.11 | 85.76±4.51 | 71.63±4.89 | 87.03±7.74 |

$$Accuracy = \frac{TP + TN}{TP + FP + TN + FN} \quad (8)$$

$$recall = \frac{TP}{TP + FN} \quad (9)$$

$$F1 - score = \frac{2 \times (recall \times precision)}{(recall + precision)} \quad (10)$$

Where,

$$precision = \frac{TP}{TP + FP} \quad (11)$$

F-1 score is the weighted average of both precision and recall and it is used in case the data is imbalanced. On the other hand, the indices 'accuracy' estimates the performance of the classifier in a better manner if there is a similar cost for FP and FN.

## IV. Results and discussions

All the experiments in this paper were implemented in Python using Keras [39] with TensorFlow [40] backend on NVIDIA P100 GPUs. We performed hyperparameter tuning to select better parameters for the proposed network as follows: We varied the number of filters for the CNN layer from {1, 2, 3, 6, 12}, the number of units for the LSTM layer from {8, 16, 32, 64}, the values of the attention length from {8, 16, 32, 64}, the length of the output of the attention layer from {6, 8, 10, 12, 14}, and the learning rate from {1e$^{-2}$,1e$^{-3}$,1e$^{-4}$, 1e$^{-5}$}. We performed 5-fold cross-validation for the samples generated from the leave-one-trial out (LOTO) strategy to select the best parameters for the datasets. To avoid bias in results, we use 10-fold cross-validation for the experiments performed on SNOW and FNOW datasets. Since the dimensions of the spatial domain (number of sensors) do not exceed 3, we limited the number of filters to 12. This was further confirmed by our experiments with a different number of filters whereby 3 filters performed the best for all datasets (except the MHEALTH dataset where we found 6 filters to give the best performance). We set the number of CNN filters to 3, number of LSTM units to 32, number of LSTM layers to 1, attention length to 32, the attention layer output length to 10, batch size to 32 and learning rate to 1e$^{-4}$. The size of the CNN filters is set be equal to the number of input filters. For MHEALTH, we set the number of CNN filters to 6. We use early stopping by monitoring validation loss.

### A. Effect of sample generation

The first experiment is intended to analyze the effect of various sample generation processes. The results are summarized in Table II. We report mean accuracy from all folds during cross-validation. For the samples generated by SNOW, the mean accuracy is significantly higher. This is evident because during data generation there is a 50% overlap in later samples, and it is more likely that some samples of training data can be mixed into the test samples. In a very similar manner, the data generated using FNOW also show high mean accuracy for most datasets. However, there is no overlap between subsequent samples during the data generation but still, there is a high chance that few data samples from the training data of the same subjects might be present in the testing samples. On the other hand, in the case of LOTO, the mean accuracy drops significantly, especially for the WISDM dataset. We have also seen inferior results from LOTO and LOSO compared to SNOW for all cases. However, we also noticed some surprising improvements in mean accuracy using LOTO for the most complex datasets UTD-MHAD1 and WHARF. For LOSO, we observed a high variance in the performance over folds, which was expected because the testing is performed on unseen subject data and therefore the performance varies over subjects. In conclusion, LOTO has no overlap in data samples, high accuracy and low variance. Henceforth, we will use LOTO for all experiments and comparative analysis.

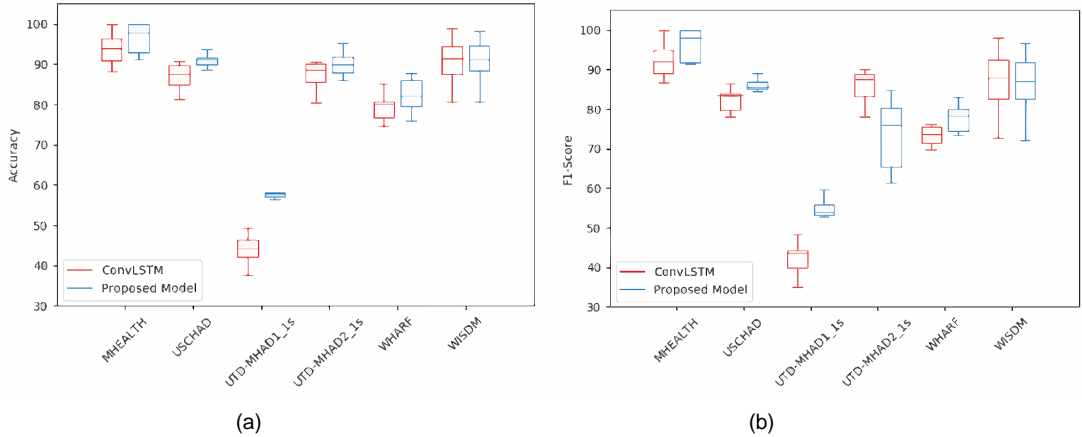

Figure 2 Box plot for comparative analysis of the proposed approach with baseline, ConvLSTM using (a) accuracy and (b) F1-score. Both indicators of performance indirectly demonstrate that the accuracy and reliability are better and stable in the proposed model than in compared one.

## B. Comparison with baseline

In the second set of experiments, we tested our proposed approach against the baseline approach ConvLSTM. We do this, because we wish to highlight the importance of including the self-attention mechanism in the proposed network. The baseline architecture resembles the DeepConvLSTM model presented in [31]. However, the DeepConvLSTM model proposed in [31] has a large number of trainable parameters which exceed the ones in our proposed approach. We therefore reduce the number of trainable parameters by modifying the number of CNN and RNN layers in the model presented in [31] for an unbiased comparison. For ConvLSTM, we set hyperparameters as follows: number of LSTM units to 32, number of LSTM layers to 1, CNN filters to 3, learning rate to 0.001. We set the initial number of epochs to 100 and introduced early stopping on validation loss with the patience of 10.

The comparative results for the proposed approach and baseline method (ConvLSTM) are given in Table III. Since the datasets are imbalanced, accuracy is insufficient for analysis and fair comparison, hence, we use recall, and F-1 scores as well. The proposed approach shows significant improvement in all datasets except MHEALTH (p-value = 0.345) and WISDM (p-value = 0.417) where the improvements are not significant. This can be attributed to the fact that the classification accuracies in both these cases are already above 90%. For the dataset UTD-MHAD1, our approach shows very high improvements (p-value = $9 \times 10^{-7}$) in the accuracy (~16%,), recall (~12%), and F1-Score (~12%). Our proposed approach shows significant improvement for USCHAD (p-value = 0.0027), UTD-MHAD2 (p-value = 0.0418), and WHARF (p-value = 0.0345). Also, the recall and F1-score higher for the ConvLSTM on UTD-MHAD2 dataset compared to our approach. In comparison to other datasets viz. MHEALTH (262), USC-HAD (840), UTD-MHAD1 (617), WHARF (884), and WISDM (402) the UTD-MHAD2 dataset has just 190 trials. This is also reflected in the low improvement in performance over the baseline in case of the MHEALTH dataset. Further, for deeper analysis, we also plotted box plots using both accuracy and F1-scores as shown in Fig. 2. Fig. 2 (a) and Fig. 2 (b) show the boxplots of the accuracies and F1-scores obtained with multiple folds under the LOTO scheme from different datasets using the proposed model and the ConvLSTM model. The proposed model performs better than the baseline ConvLSTM model in terms of both F1-score and accuracies except in cases where the number of trials is relatively low. In cases where the number of trials is low (for example USC-MHAD2 or MHEALTH), the training set of the DNN is small and the performance is similar to the baseline. Overall, the proposed approach shows significant improvement compared to ConvLSTM.

## C. Comparative analysis with existing work

We compare our approach with existing state-of-the-art methods. We compare our method with approaches that use handcrafted features on traditional machine learning [16], recently developed ensemble learning with handcrafted features [7]. We also compared our method with CNN[8], [21] and CNN with LSTM (ConvLSTM) [31] which use automatic feature generation. Since a direct comparison is not possible, due to high variability in hyperparameters, dataset types, and data-generation techniques, we perform all experiments on our prepared datasets.

Since, the dataset and its preparation are different in the current work from the original work to compare, the network architecture, and optimal hypermeters will also vary. Therefore, in the following sections, we first describe network architecture and hyperparameter settings used in our experiments and then we compare the results of the proposed architecture with the existing state-of-the-art approaches.

For [16], we extracted 5 statistical features, such as mean, standard deviation, average resultant acceleration, and time between peaks from the time series signals from all sensors of the given datasets. For classification we use multilayer perceptron with solver set to lbfgs, alpha set to $10^{-5}$, learning rate of 0.3 and momentum of 0.2. Number of hidden layers are set by (number of features + number of classes)/2.

Table IV Comparative analysis of the proposed approached in terms of accuracy with existing work in literature.

| Ref | classifier | features | MHEALTH | USC-HAD | UTD-MHAD1 | UTD-MHAD2 | WHARF | WISDM |
|---|---|---|---|---|---|---|---|---|
| [16] | MLP | H | 89.49±7.39 | 77.95±3.04 | 14.5±3.04 | 70.91±7.2 | 51.71±13.31 | 78.26±7.76 |
| [7] | Ensemble | H | 91.72±6.94 | *87.35±2.48* | *47.63±3.27* | 82.5±3.4 | 61.14±2.37 | 79.97±5.69 |
| [8] | 2D-CNN | self | 89.9±6.31 | 86.22±2.29 | 47.02±2.41 | 71.48±5.4 | 75.88±3.73 | 86.67±6.44 |
| [31] | CNN, LSTM | --do-- | *93.80±3.87* | 87.05±3.2 | 43.92±3.71 | *87.06±3.74* | *77.80±4.07* | *89.83±7.04* |
| Proposed | Self-attention | --do-- | **94.86±7.65** | **90.88±1.47** | **58.02±2.29** | **89.84±2.97** | **82.39±3.93** | **90.41±5.59** |

For [7], features are generated as in [16]. We ensemble j48 (java implementation of C4.5 algorithms), MPL, and logistic regression. The parameter settings for MLP unit is same as in [16]. For j48 (decision tree) and logistic regression, we use the scikit-learn Python library [41] with default parameter settings. For ensemble, we use *voting classifier* from scikit-learn with '*soft*' voting.

For comparison with CNN, we use two references[8], [21]. Since, in [21] the CNN network was too shallow and we could not get good results, we choose to go with only[8]. For[8], we use three layers of 2D CNN (number of filters = 18, 36, and 24 and kernel sizes = (12,2), (13,1), and (12,1) for subsequent layers) following with max-pooling (pool size 2,1). We use

'ReLU' at convolutional layers while 'softmax' at output layer with 'adadelta' optimizer. For other hyperparameters, we use default parameters in Keras with Tensorflow as backend. However, for the datasets WHARF and WISDM, the sampling rate was too small, therefore, it was not possible to use large convolutional kernels. Therefore, we chose kernels with sizes (3, 2), (3, 1) and (3, 1), and other settings were the same as for other datasets.

We summarize the results in terms of accuracy in Table IV. For clarity, we have featured the best performer with bold and second-best with italics and underlines. In terms of accuracy, our method is superior to all methods found in the literature. The improvement obtained by the proposed method over the previous state-of-the-art is statistically significant for all datasets (except WISDM). The lowest performance is achieved using MLP with handcrafted features on all datasets. However, on the statistical features, we see an improvement in accuracy for all datasets than MLP when we use ensemble learning (j48, MLP, and logistic regression), especially for UTD-MHAD1. We do not see any significant difference between ensemble learning and 2D CNN and we found mixed results for different datasets.

## V. CONCLUSIONS

In this paper, we proposed a deep learning architecture that considers the spatio-temporal aspects of data from multiple wearable sensors and uses self-attention to identify embedding combinations for decoding human activity. While the CNN and RNN layers encode the spatial and temporal features, respectively, the self-attention layer generates feature representation from the embeddings of sensor time-series generated by the CNN and RNN. We show that the proposed architecture gives comparable performance for different sample generation schemes on the MHEALTH, USC-HAD, UTD-MHAD, WHARF and WISDM datasets. On comparing the performance of the proposed architecture with a baseline based on CNN and RNN models, and with previous state-of-the-art approaches, we find that the proposed approach leads to an improvement in accuracy for activity recognition. Thus, the proposed architecture has wide applications in human activity recognition and can be scaled for a large number of sensors and time points. An extension of the present work would involve trying out different attention mechanisms including the global and local attention [33] and comparing the performance of self-attention mechanism for human activity recognition.